\documentstyle[12pt]{article}
\def\bfl{\begin{flushleft}}
\def\efl{\end{flushleft}}
\def\bfr{\begin{flushright}}
\def\efr{\end{flushright}}
\def\bc{\begin{center}}
\def\ec{\end{center}}
\def\be{\begin{equation}}
\def\ee{\end{equation}}
\def\ba{\begin{eqnarray}}
\def\ea{\end{eqnarray}}
\def\nn{\nonumber }
\def\lb#1{\label{#1}}

\def\text#1{\mbox{#1}}

\def\drm{\text{d}}

\def\Sign#1{\, \text{sign}\left[#1\right] }

\def\Po{Poincar\'e}
\def\RN{Reissner-Nordstr\"om~}
\def\Kummer#1#2#3{\, \text{M}\left(#1,\,#2;\,#3 \right) }

\begin{document}
\bc
~~\\
\bfr
Int. J. Mod. Phys. D 8 (1999) 165-176\\
LANL e-print gr-qc/9807012
\efr
~~\\
{\large \bf
Extended particle models based on hollow singular 
hypersurfaces in general relativity: Classical and quantum aspects
of charged textures
}

~~\\
Konstantin G. Zloshchastiev\\
~~\\
http://zloshchastiev.webjump.com\\
%www.geocities.com/zlosh\\
~\\
Received: {\it 6 July 1998 (LANL); 12 November 1998 (IJMPD)}

\ec

~~\\

\abstract{
In present paper we construct classical 
and quantum models of an extended charged particle.
One shows that consecutive modelling can be based on the hollow 
thin-wall charged texture 
(in the hydrodynamical approach of a perfect fluid) 
which
acquires gravitational mass due to Einstein-Maxwell interaction.
We demonstrate that such a model has equilibrium states
at the radius equal to the established 
classical radius of a charged particle.
Also we consider quantum aspects of the theory and
obtain the (internal) Dirac sea conception in a natural way. 
Besides, the phenomenological unification on the mass level of 
the two families of elementary 
particles, charged pions and electrons and positrons, evidently arises 
as the effect induced by classical and quantum gravity prior to
Standard Model.
Finally, in the cosmological connection our model proposes the answer on 
the important question, what are the real sources of texture matter. Besides, 
the texture hypothesis means that in the early Universe the 
topological texture foam phase existed before the lepton-hadron one.

}

~\\

PACS number(s):  04.40.Nr, 04.60.Ds, 11.27.+d\\

~~\\

\newpage
%\large

\section{Introduction}\lb{s-in}

The very important question, what is the origin of the observable properties
of the known elementary particles, such as charge and 
mass\footnote{The 
fundamental properties of the charged particles, (a) the electron: mass 
$m_e = 0.511~ \text{MeV} = 6.8 \times 10^{-58}\text{m}$,
charge $Q_e = 1.38 \times 10^{-36}\text{m}$,
$m_e/Q_e=4.9 \times 10^{-22}$,
classical radius $R_e = Q_e^2/m_e = 2.8 \times 10^{-15}\text{m}$;
(b) the $\pi^\pm$- meson: mass 
$m_{\pi} = 139.6~ \text{MeV} = 1.86 \times 10^{-55}\text{m}$,
charge $Q_{\pi} = Q_e$,
$m_{\pi}/Q_{\pi}= 1.35 \times 10^{-19}$,
classical radius $R_{\pi} = 10^{-17}\text{m}$.},
seems to be resolved within the frameworks of the theory considering these
particles as extended entities.

The first model of an extended particle was proposed about a hundred 
years ago by Lorentz \cite{roh}.
He suggested a model of the electron as the body having pure 
charge and no matter.
His model appears to be highly unstable, the electric repulsion
should eventually lead to explosion of the configuration.
This circumstance was corrected by \Po~who introduced the stresses
to provide stability of the Lorentz model \cite{poi}.
In general relativity the Lorentz-\Po's~ideas have gained 
new sounding in terms of repulsive gravitation \cite{gro}, 
and interest to such models has been renewed, especially in connection
with the modelling of exotic extended particles.
Since that time many modifications of them have been done.
In spite of the fact that majority of the proposed models  seems to be 
appropriate for a charged 
spinless particle (e.g., $\pi^\pm$-meson) rather then for the electron 
(except the works \cite{il}), they nevertheless contain some consistent 
descriptions and concepts concerning an electron when assuming that presence
of spin insufficiently violates spherical symmetry \cite{zlo008}.
The main idea of the most popular electron models is to consider the 
matching of the (external) \RN solution of the Einstein-Maxwell field 
equations,
\be
\drm s^2 =- 
\left(1-\frac{2 M}{r} + \frac{Q^2}{r^2} \right) \drm t^2 + 
\left(1-\frac{2 M}{r} + \frac{Q^2}{r^2} \right)^{-1} \drm r^2 +
r^2 \drm \Omega^2                                                \lb{eq1}  
\ee
(we will work in terms of the gravitational units), and some
internal solution; thereby this matching is performed
across the boundary surface (which is considered as a particle surface), 
i.e., the first and second quadratic forms are continuous on it,
that appears to be the well-known Lichnerowicz-Darmois junction 
conditions \cite{gau}.
At the same time, more deep investigations of the electromagnetic 
mass theories, based
on this junction, elicit a lot of the features which point out
some imperfection of the particle models based on the 
boundary surfaces (i.e., on the discontinuities of first kind).

The first of them is the accumulation of electric charge on a core
boundary.
Let us show it for the model \cite{cc},
where the external \RN solution (\ref{eq1}), at 
$r > R$, $R$ is a core radius, is matched with the de Sitter one, 
at $r\leq R$,
\be
\drm s^2 =- 
(1-\lambda^2 r^2)\, \drm t^2 + 
(1-\lambda^2 r^2)^{-1} \drm r^2 +
r^2 \drm \Omega^2,                                                 \lb{eq2}  
\ee
$\lambda^2 = 8 \pi \varepsilon_v/3$, thereby they obtained 
$M = (4 \pi \varepsilon_v/3) R^3 + Q^2/2 R$, 
$\varepsilon_v$ is the physical vacuum energy density.
The stress-energy tensor components for this solution look like those
for the polarized vacuum \cite{zel}:
\ba
&& T^0_0 = \varepsilon_v = ~^{(3)}\!\rho + Q^2(r)/8 \pi r^4, \nn\\
&& T^1_1 = \varepsilon_v = -~^{(3)}\!p + Q^2(r)/8 \pi r^4, \lb{eq3}\\
&& T^2_2 =T^3_3 = 
\varepsilon_v = - ~^{(3)}\!p_\bot - Q^2(r)/8 \pi r^4, \nn
\ea
where $Q(r)$, $^{(3)}\!\rho$, $^{(3)}\!p$ are the charge, energy density,
and pressure inside $r$, respectively.
The generalization of the Tolman-Oppenheimer-Volkov equation to this case
yields \cite{trk}
\be
\frac{\drm }{\drm r} ~^{(3)}\!p =  
\frac{1}{8 \pi r^4} \frac{\drm }{\drm r} Q^2(r)
\quad
r\leq R,                                                          \lb{eq4}
\ee
hence
\be
Q(r) = Q\, \theta (r - R),                                          \lb{eq5}
\ee
where $\theta (r)$ is the Heaviside step function.
Thus, there is no electric field inside the core, and charge is 
accumulated on the boundary only.
However, the discontinuities of first kind cannot have surface charge 
distribution {\it a priori}.
The second feature is the energy of the physical vacuum is transferred 
to a boundary as well \cite{col,cg}.
Nevertheless, the discontinuities of first kind can not have 
surface energy density too.

\section{Singular shell approach}\lb{s-si}

The above-mentioned facts say that we must consider the system of 
the Einstein-Maxwell field equations plus singular (infinitely thin) 
shell \cite{dau,isr,mtw} instead of the simple boundary surface.
Such a shell is the discontinuity of second kind 
(the first quadratic form
is continuous across it but the second one has a finite jump), and,
unlike the boundary surface, can have the proper surface charge $Q$ and
the stress-energy tensor
\be
S_{ab}=\sigma u_a u_b + p (u_a u_b +~ ^{(3)}\!g_{ab}),            \lb{eq6}
\ee
where $\sigma$ and $p$ are respectively the surface energy density and 
pressure, $u^a$ is the timelike unit tangent vector, 
$^{(3)}\!g_{ab}$ is the 3-metric on a shell.

Thus, let us study the spherical singular shell inducing the 
external \RN spacetime $\Sigma^+$ (\ref{eq1}). 
Keeping in mind the second feature pointed out above we require that
the internal spacetime $\Sigma^-$ has to be the Minkowski flat one. 
The metric of the (2+1)-dimensional spacetime $\Sigma$ 
of the shell can be written in 
terms of the shell's proper time $\tau$ as
\be
^{(3)}\!\drm s^2 = - \drm \tau^2 + R^2 \drm \Omega^2,             \lb{eq7}
\ee
where $R(\tau)$ is the proper radius of the shell.

Then the (3+1)-split Einstein-Maxwell field equations give the
two groups of equations.
The first one is the Einstein-Maxwell equations on the shell, integrability
conditions of which yield both the conservation law of shell's matter
\be
\drm \left( \sigma ~^{(3)}\!g \right) +
p~ \drm \left( ~^{(3)}\!g \right) 
+ ~^{(3)}\!g~  \Delta T^{\tau n}\, \drm \tau =0,                   \lb{eq8}
\ee
where 
$^{(3)}\!g=\sqrt{- \det{(^{(3)}\!g_{ab})}} = R^2 \sin{\theta}$,
$\Delta T^{\tau n} = (T^{\tau n})^+ - (T^{\tau n})^-$,  
$T^{\tau n}=T^{\alpha\beta} u_\alpha n_\beta$ is the
projection of the stress-energy tensors in the $\Sigma^\pm$
spacetimes on the tangent and normal vectors 
(it can immediately be verified that $T^{\tau n} \equiv 0$ for the 
Schwarzschild, \RN, de Sitter and Minkowski spacetimes), 
and the electric charge conservation law which in our case can be reduced 
to the relation
$
Q=\text{constant}.
$
The second group includes the equations of motion of a shell which are 
the Lichnerowicz-Darmois-Israel junction conditions 
\be
(K^a_b)^+ - (K^a_b)^- = 4 \pi\sigma (2 u^a u_b + \delta^a_b),
                                                                    \lb{eq9}
\ee
where $(K^a_b)^\pm$ are the extrinsic curvatures for the spacetimes 
$\Sigma^\pm$ respectively.
As for the electromagnetic potential it is zero inside the shell, has a 
jump across the shell and turns to be 
equal to the Coulomb's one outside the shell.
Taking into account eqs. (\ref{eq1}), (\ref{eq7}), and (\ref{eq8}),
the $\theta\theta$ and $\tau\tau$ components of eq. (\ref{eq9}) yield
\be
\epsilon_+ \sqrt{\dot R^2 + 1 - \frac{2 M}{R}+\frac{Q^2}{R^2}}
-
\epsilon_- \sqrt{\dot R^2 + 1} =
-\frac{m}{R},                                            \lb{eq10}
\ee
\be
\frac{\ddot R + M/R^2 -Q^2/R^3}
     {
      \epsilon_+ \sqrt{\dot R^2 + 1 - \frac{2 M}{R}+\frac{Q^2}{R^2}}
     }
-
\frac{\ddot R}
     {
      \epsilon_- \sqrt{\dot R^2 + 1}
     }
=-\frac{\drm}{\drm R} \left( \frac{m}{R} \right),              \lb{eq11}
\ee
where $\epsilon_\pm = \Sign{(K_{\theta\theta})^\pm}$, 
$m=4\pi\sigma R^2$ is interpreted as the (effective) bare mass of the shell,
$\dot R = \drm R/\drm \tau$ etc.
The root sign $\epsilon = +1$ if $R$ increases in 
the outward normal of the shell, and $\epsilon = -1$ if $R$ decreases.
Thus, only under the additional condition $\epsilon_+ = \epsilon_-=1$ we 
have the ordinary shell.
Below we will deal with such shells only.

In addition, independently of the Einstein equations,
the equation of state for matter on the shell $p=p(\sigma)$
should be added as well.
The simplest equation of state we choose is the linear one of a barotropic 
fluid
\be
p=\eta \sigma,                                                  \lb{eq12}
\ee
including the most physically suitable cases, e.g., the dust ($\eta = 0$), 
radiation fluid
($\eta = 1/2$) and bubble matter ($\eta = - 1$).
Then the constant $\eta$ remains to be arbitrary and will be determined 
below from the equilibrium conditions.
It can easily be seen that eqs. (\ref{eq8}), (\ref{eq10}) and 
(\ref{eq12}) form a complete system.
Solving equations (\ref{eq8}) and (\ref{eq12}) together, we obtain
\be
\sigma =\frac{\alpha}{4 \pi} R^{-2(\eta+1)},                    \lb{eq13}
\ee
where $\alpha$ is the integration constant 
(having the dimension $\text{cm}^{2\eta+1}$ in geometrical units) 
related to the surface energy 
density at some fixed $R$, $\alpha>0$ for ordinary shells.
Then 
\be
m = \alpha R^{-2 \eta},                                        \lb{eq14}
\ee
and eqs. (\ref{eq10}), (\ref{eq11}) can be rewritten more strictly:
\ba
&&\sqrt{\dot R^2 + 1 - \frac{2 M}{R}+\frac{Q^2}{R^2}}
- \sqrt{\dot R^2 + 1} =
-\frac{\alpha}{R^{2\eta+1}},                                \lb{eq15}\\
&&\frac{\ddot R + M/ R^2 -Q^2/R^3 }
     {
      \sqrt{\dot R^2 + 1 - \frac{2 M}{R}+\frac{Q^2}{R^2}}
     }
-
\frac{\ddot R}
     {
      \sqrt{\dot R^2 + 1}
     }
=\frac{\alpha (2\eta +1)}{R^{2(\eta+1)}}.                      \lb{eq16}
\ea
Now we consider this system in the equilibrium $\dot R=\ddot R=0$ at
certain $R=R_{\text{eq}}$.
Then these equations turn to be the equilibrium conditions \cite{fhk}
\ba
&&\sqrt{1 - \frac{2 M}{R_{\text{eq}}}+\frac{Q^2}{R_{\text{eq}}^2}}
=
1 -\frac{\alpha_{\text{eq}}}{R_{\text{eq}}^{2\eta+1}},           \lb{eq17}\\
&&\frac{M R_{\text{eq}} -Q^2}
     {
      \sqrt{1 - \frac{2 M}{R_{\text{eq}}}+\frac{Q^2}{R_{\text{eq}}^2}}
     }
=\frac{\alpha_{\text{eq}} (2\eta +1)}{R_{\text{eq}}^{2\eta-1}}.    \lb{eq18}
\ea
Further, it is well-known that the classical radius $R_c$ of a charged
particle can be defined as the radius at which 
the metric $1-2M/R+Q^2/R^2$ 
approaches a minimum hence $R_c = Q^2/M$ (see footnote 1). 
If one assumes $R_{\text{eq}} = R_c$ the last equation yields
\be
\eta =- \frac{1}{2} \Rightarrow \sigma =  \frac{\alpha}{4\pi R}.      
                                                               \lb{eq19}
\ee
The surface energy density determined by this expression appears to be
the 2D counterpart of the time-time component of the cosmological texture
stress-energy tensor 
(see, e.g., Eq. (3.1) in the paper by Davis \cite{tur}) if
one takes into account the reduction of dimensionality.
This is an expected result: from the viewpoint of the 2D observer
``living'' on the shell it seems for him to be the whole universe with
the scale factor $R$.
In this connection the {\it dimensionless} 
integration constant $\alpha$ obtains the
sense of the length of the quadruplet vector from the model giving rise
to the breaking of a global symmetry and thus to the appearence of
textures.

As for eq. (\ref{eq17}) then considering eq. (\ref{eq19}) 
we obtain
\be
\alpha_{\text{eq}}=1-\sqrt{1 - (M/Q)^2}.                       \lb{eq20}
\ee
Hence it follows $M<Q$, $0 < \alpha_{\text{eq}} <1$.
For the known particles the value $M/Q$ turns to be small, 
therefore
\be
\alpha_{\text{eq}} = \frac{1}{2} \frac{M^2}{Q^2} 
+ O \left( \frac{M^4}{Q^4} \right).                            \lb{eq21}
\ee
Therefore, from eq. (\ref{eq19}) it follows that the 
equation of state of matter on the shell 
\be
\sigma + 2 p=0                                                 \lb{eq22}
\ee
has to be the necessary 
condition of equilibrium of the hollow charged shell.
The three-dimensional analogue of this equation is the equation of state
of the global texture \cite{tur,dad},
\be
^{(3)}\!\varepsilon  + 3~^{(3)}\!p=0,                          \lb{eq23}
\ee
characterizing vacuum 
energy or the new kind of exotic matter called k-matter \cite{vil}.
Thus, our hollow charged singular model should be based on the 
two-dimensional texture fluid to have equilibrium states. Of 
course, we assume here at least the two physical approximations: 
that of thin 
layer of matter (i.e. 3D $\approx$ 2D), and the approximation of fluid 
hence
we suppose that texture topological defects are coupled to each other and 
(self)interact with gravity only through their macroscopical equation 
of state.

The gravitational mass of the neutral texture, as was expected, 
has to be zero in equilibrium, just assume $Q=0$ in (\ref{eq18}).
As for the charged texture, then
from eq. (\ref{eq18}), taking into account (\ref{eq19}), one can see that 
it acquires the gravitational mass due to the electromagnetic interaction
thus we have another electromagnetic mass model. 

Finally, it should be noted that there exists a direct connection of
textures with the other topological defects, namely, 
global monopoles \cite{dad,cg,zlo}.
In the present paper such a connection arises in a natural way when
studying quantum dynamics of the model.

\section{Quantum dynamics}\lb{s-qu}

Considering eq. (\ref{eq19}), the equation of motion of a
two-dimensional charged texture (\ref{eq10}) can be rewritten as
a conservation law of total mass-energy for the charged
relativistic particle
of variable rest mass
\be
M=m\sqrt{1+\dot R^2} + \frac{Q^2-m^2}{2 R},              \lb{eq24}
\ee
where
\be
m= \alpha R.                                              \lb{eq25}
\ee
If one works in terms of the flat time $T$ inside the shell,\footnote{To 
a high accuracy the spacetime $\Sigma_+$ for the distant external observer 
can be supposed flat as well.} 
$\dot T^2 - \dot R^2 = 1$, then eq. (\ref{eq24}) reads
\be
M= \frac{m}{\sqrt{1 - R_T^2}} + \frac{Q^2-m^2}{2 R},              \lb{eq26}
\ee
where $R_T = \drm R /\drm T$.
Squaring this expression, introducing the momentum
$\Pi = m \dot R = m R_T/\sqrt{1-R_T^2}$, and taking into account
eq. (\ref{eq25}), we obtain the energy-momentum conservation law of a 
(1+1)-dimensional relativistic point particle:
\be
\left(
       M - \frac{Q^2-\alpha^2 R^2}{2 R}
\right)^2 - \Pi^2 = \alpha^2 R^2.                                 \lb{eq27}
\ee
There exist several approaches to the quantization of spherically
symmetric thin shells in general relativity,
that is connected with the different ways of constructing the Hamilton 
formalism and choice of gauge conditions \cite{hb,hkk,bkkt,not}.
Below we perform the direct quantization of the conservation law 
(\ref{eq27}), thereby the ideology of such quantization coincides 
with that of ref. \cite{hkk}
due to the presence of flat spacetime inside the shell.
The last method appears to be in evident agreement with the correspondence 
principle, and
we can consider the two-dimensional charged texture as the
stationary quantum system with a 
single radial degree of freedom, in contradistinction to the thick layer of 
matter which has an infinite number of degrees of freedom.

Substituting the operator $\hat\Pi=-i\partial_R$ for the momentum $\Pi$
(we use Planckian units), 
we consider the singular Stourm-Liouville problem for the self-adjoint 
operator given by the constraint (\ref{eq27}) and obtain the equation 
for the spatial wave function $\Psi (R)$  of a shell:
\be                                                         \label{eq28}
\Psi^{\prime\prime} +
\left[
      M^2 - \frac{\alpha^2 Q^2}{2} - \frac{M Q^2}{R} 
      + \frac{Q^4}{4 R^2}
      + \alpha^2 M R - \alpha^2 R^2 (1-\alpha^2/4)
\right]\Psi =0,
\ee
which describes quantum oscillations of a particle size.
It should be noted that this equation could be ruled out also from
the minisuperspace approach \cite{vil2} 
(requiring no time slicing of the basic manifold) 
if we suppose the Lagrangian
\[
L = \frac{m \dot R^2}{2} 
+ \frac{R}{8\alpha}
\left(
       \alpha^2 + \frac{2M}{R} - \frac{Q^2}{R^2}
\right)^2
- \frac{\alpha R}{2}.
\]
Indeed, the equation of motion, following from it, being 
once integrated yields the conservation law (\ref{eq27}) up to the
integration constant which has sense of the super-Hamiltonian and can
be supposed vanishing in spirit of the Wheeler-DeWitt's approach.
Then (\ref{eq28}) could be obtained as the corresponding
Wheeler-DeWitt equation when performing the canonical quantization
of the model.

Further, equation
(\ref{eq28}) regrettably can not be resolved in known functions.
Instead of it we study the two asymptotical wave equations following from 
it.
Indeed, there exists the boundary value of $R$, 
$R_b = Q/\alpha \in (Q,+\infty)$, such that:
at $R \gg R_b$ one can neglect the third and fourth terms in square 
brackets, and obtain
\be                                                         \label{eq29}
\Psi^{\prime\prime} +
\left[
      M^2 - \frac{\alpha^2 Q^2}{2} 
      + \alpha^2 M R - \alpha^2 R^2 (1-\alpha^2/4)
\right]\Psi =0,
\ee
at $R \ll R_b$ the last two terms turn to be negligibly small, therefore,
\be                                                         \label{eq30}
\Psi^{\prime\prime} +
\left[
      M^2 - \frac{\alpha^2 Q^2}{2} - \frac{M Q^2}{R} 
      + \frac{Q^4}{4 R^2}
\right]\Psi =0.
\ee
Below we consider these cases separately.

(a) $R \gg R_b$.
Then the solution of eq. (\ref{eq29}), which is regular in the chosen 
applicable domain of half-line type, can be expressed by means of the 
Kummer confluent hypergeometric function $\Kummer{a}{b}{z}$ \cite{as}:
\be                                                         \lb{eq31}
\Psi (z) = 
\exp{
     \left[
              -\frac{z^2 }{ 8 \alpha^3 \beta}
     \right]
    }           z 
\Kummer{\lambda}{\frac{3}{2}}{\frac{z^2 }{4 \alpha^{3}\beta}},
\ee
where
\ba
&&\beta = (1-\alpha^2/4)^{3/2}, \nn\\
&&z/ \alpha^2 = 2 (1-\alpha^2/4) R - M, \nn\\
&&4\lambda = 3-
     \frac{
                2 M^2 - \alpha^2 Q^2 (1-\alpha^2/4)
           }
           {
            2 \alpha \beta
           }.                                                \nn
\ea
Let us find the bound states of this system, for which
we require the solution (\ref{eq31}) to be asymptotically vanishing at
$R \rightarrow +\infty$.
It is satisfied on the poles of the gamma-function 
$\Gamma (\lambda)$ \cite{zlo}:
\[
\lambda=-n,
\]
where $n$ is a non-negative integer.
From this expression we obtain the total mass spectrum:
\be                                                   \lb{eq32}
M_n = \alpha
\sqrt{1-\alpha^2/4}
\sqrt{
     \frac{\sqrt{1-\alpha^2/4}}{\alpha}
     (4 n +3) 
     + \frac{Q^2}{2}
     }.
\ee
It should be noted that the value $R_b\sim Q^3/M^2$ (\ref{eq21})
appears to be huge for the known elementary particles: taking into 
account the estimation (\ref{eq21}) we obtain that for the electron
$R_b\sim 10^{42}$ ($10^9$ cm), for the $\pi^\pm$-mesons $R_b\sim 10^3$ cm.
Thus, the wave equation (\ref{eq29}) describes either highly excited mass 
levels of the known particles or some massive exotic particle.
In this connection the next limit case, $R\ll R_b$, seems to be
more appropriate for the modelling of  charged elementary particles.

(b) $R \ll R_b$.
The equation (\ref{eq30}) up to redefining of the constants $Q$ and $\alpha$
coincides with the quantum equation for the other topological defect,
hollow monopole, which was studied as a private case in the 
work \cite{zlo}.
Then the solutions of eq. (\ref{eq30}) vanishing at zero are the functions
\be                                                          \label{eq33}
\Psi_\pm (Q,~M;~R>0) = 
\text{e}^{\pm i \sqrt{\chi} R} 
R^{\lambda_\zeta}
\Kummer{
        \lambda_\zeta -
        \frac{M Q^2}{2 \sqrt{- \chi}
                    }
       }
       {2 \lambda_\zeta}
       {
        2 i \sqrt{\chi } R
       },
\ee
where
\[
\chi =   M^2- \frac{\alpha^2 Q^2}{2},
\]
\be                                                         \label{eq34}
\lambda_\zeta =\frac{1+\zeta \sqrt{1-Q^4}}{2},                      
\ee
the root sign $\zeta=\pm 1$ denotes the additional splitting of the solutions
(for more details and discussion see ref. \cite{hkk}).
It should be noted that such a splitting arises both in relativistic and
non-relativistic quantum mechanics of the electron in the hydrogen atom.
However there the solutions of wave equations corresponding to $\zeta=-1$
were irregular and thus did not satisfy the quantum 
boundary condition at the origin.
In present case the both cases $\zeta=\pm 1$ satisfy the quantum
boundary conditions as it can be checked, thus we can not make a justified 
choice between them.  
Moreover, it will be shown below that
the physical sense of $\zeta$ is very important.
The necessary condition for the existence of 
quantum bound states on an infinite applicable domain 
is the vanishing of the eigenfunctions both at the origin and at infinity,
thus we extrapolate quantum boundary to infinity that corresponds to the 
approximation $\alpha \ll 1$ as well.
The first condition is already satisfied by the choice of the solution 
(\ref{eq33}).
The second will be satisfied again on the gamma-function poles, i.e., at
\be                                                          \label{eq35}
\lambda_\zeta-
        \frac{M Q^2}{2 \sqrt{-\chi}
                    } = -n,
\ee
hence we obtain the total mass spectrum
\be                                                          \label{eq36}
M_n =\pm \frac{\alpha Q}{\sqrt{2}}
      \left[
            1+Q^4 \left(
                       {\cal N} + \zeta \sqrt{1-Q^4}
                  \right)^{-2}
      \right]^{-1/2},
\ee
for the ground state $n=0$ this equation yields
\be                                                          \label{eq37}
M_0 =\pm \frac{\alpha Q}{2}
      \sqrt{
            1+ \zeta \sqrt{1-Q^4}
           },
\ee 
where ${\cal N} = 2 n +1 = 1, 3, 5,...$ (compare with ref. \cite{zlo}).
In these expressions
$\alpha$ remains to be a free dimensionless parameter of topological 
nature (\ref{eq19}),
the root signs ``$\pm$'' correspond to upper and 
lower energy continua and they are directly connected neither with 
$\zeta$ and $\lambda_\zeta$ nor with parameters of spacetimes outside 
and inside the shell.
However we can associate them with the sign of the electric charge $Q$, 
i.e., they appear to be discriminating the particles and antiparticles  
in spirit of the Dirac sea conception 
(even without use of spinors though from 
(\ref{eq27}) one can obtain also the 
(1+1)-dimensional spinor equation in the same manner as the Dirac 
equation was deduced from the Klein-Gordon-Fock one \cite{gz}).
It is interesting to note that this interpretation was impossible for
the charged dust hollow shells \cite{zlo} despite the energy continua
exist for them as well.
Thus, it is another case for the texture model (\ref{eq22}).

Further, from expressions (\ref{eq33}) - (\ref{eq35}) it follows that the 
necessary conditions of the existence of bound states are the inequalities
\ba
&& M^2 < \alpha^2 Q^2/2,                           \label{eq38}\\ 
&&|Q| \leq 1.                                                  \label{eq39}
\ea

The first is the condition of energy ellipticity, which is usual for 
the existence of bound states in quantum mechanics.
The second inequality determines the extremal value of $Q$: 
$Q_{\text{~limit}} \approx \sqrt{137}\, Q_e$.
If it is not satisfied, a quantum instability arises.
Similar situation takes place in quantum 
electrodynamics at the description of a relativistic electron in a static 
field of the hypothetical nucleus with charge $Z > 137$ \cite{bd}.

Let us find a physical sense of $\zeta$.
To do it, following eq. (\ref{eq37}) we calculate the ratio of the 
ground state masses
\be \lb{eq40}
\frac{M_0^{(\zeta=+1)}}{M_0^{(\zeta=-1)}} = \frac{1+\sqrt{1-Q^4}}{Q^2}
\ee 
for the elementary charge $Q_e \approx 1/\sqrt{137}$.
We obtain that this quotient,
\be \lb{eq41}
\frac{M_0^{(\zeta=+1)}}{M_0^{(\zeta=-1)}} \approx 273.9,
\ee
almost coincides with that for the charged pions and 
electron (see footnote 1),
\be                                                                \lb{eq42}
\frac{m_\pi}{m_e} \approx 273.2.
\ee
Therefore, we see that
\be                                                                \lb{eq43}
\frac{M_0^{(\zeta=+1)}}{M_0^{(\zeta=-1)}} = 
\frac{m_\pi}{m_e}
\ee
to a high accuracy ($\sim 0.3\, \% $), and hence we can suppose that the 
sign $\zeta=+1$ corresponds to the family of charged mesons $\pi^\pm$, 
whereas
the sign $\zeta= -1$ corresponds to the electrons and positrons $e^\pm$.
Thus, the quantum equation (\ref{eq30}) and 
mass spectra (\ref{eq36}), (\ref{eq37})
describe the four charged elementary particles, $\pi^\pm$ and $e^\pm$
(at least, in ground states).
This result is consistent with the fact pointed out above that 
relativistic spherically symmetric models are the most appropriate
for these particles among the rest (i.e., they satisfy well enough with 
the assumption of spherical 
symmetry both on the classical and quantum levels of modelling).
Of course, such a unification remains to be only phenomenological  
(because of we consider only gravitational effects): families
of these particles are remote from each other within the frameworks of
recent Standard Model.
Nevertheless, it was shown that following our 
geometrodynamical extended model the self-gravitating equilibrium 
spherically symmetric distribution of the hypothetical 
texture matter, at first, on the classical level has the radius equal to 
the long established classical one for a charged particle, 
and, at second, on the quantum level reproduces 
the (internal) Dirac sea, and the known mass spectra follow 
naturally from the only spectrum (\ref{eq36}).

Therefore, the physical picture seems to be as follows.
The charged particle parameters such as energy spectra 
(discrete or continuous) 
are obtained within the frameworks of 
relativistic quantum electrodynamics either
from Dirac or Klein-Gordon-Fock equations (in dependence on spin)
plus the standard loop radiational and vacuum corrections depending
also on other internal degrees of freedom (hypercharges etc.), 
and explicitly contain
the particle gravitational mass as a free parameter which is traditionally 
determined from an experiment.
However, in our thin-wall 
model this mass is obtained theoretically (up to the
free multiplicative dimensionless parameter of topological nature) 
through the considering of
classical and quantum binding effects of gravitational origin, 
and this model suggests the following particle structure:
to increase the stability of a charged particle it is more preferable that 
the charged texture foam (i.e., the topological 
texture fluid containing the carriers of electrical charge which
are ``more'' elementary than electrons) is displaced onto 
particle's ``surface''\footnote{Generally speaking, in nature the 
singular surface (i.e., the boundary layer with surface tension and 
energy) 
appears everywhere where the equilibrium of configuration is provided 
by the equilibrium of opposite forces, thereby on the surface 
it takes place some concentration of matter, energy, charge, etc., 
which can be not characteristic for a whole configuration.
The considering of quantum uncertainty leads to the surface becomes
probabilistically fuzzy, and its physical characteristics 
must be understood as averaged values.}, 
see top of the paper, and its non-linear 
interaction with spacetime metric determines mass (or part of the total
mass) whereas the interior of
a particle is governed by other interactions (e.g., for charged pions it is
the $u\bar d/ \bar u d$ quark-antiquark interaction).

As for the electron's interior it is hard to say something definite now
... 
maybe the electron is ``empty'' in the sense that its interior does not
affect on its known properties almost, and in this connection one can say
that the texture matter is purely the subelectron matter.
Another question which is hard for explanation now is why the pions
as quark-antiquark combinations admit the approximation of empty interior
similarly to electron.
One can propose the three possible explanations, 
(i) the field of surface texture matter confines the quarks thereby their
charge distributions have peaks on surface and the $u\bar d/ \bar u d$ 
interaction is such that the resulting energy curves interior spacetime
very slightly in comparison with surface effects, 
(ii{\it a}) the mass spectrum $(\zeta=+1)$ is that of the pair of
unknown charged particles rather than pions, these
hypothetical particles have the same charge as pions 
but they are insufficiently more massive, $\sim 0.3 \%$,
and they have either $0$ (preferably) or $1/2$ spin,
(ii{\it b}) the mass spectrum $(\zeta=-1)$ is that of the pair of
unknown charged particles rather than electrons, these
hypothetical particles have the same charge as electrons 
but they are insufficiently less massive, $\sim 0.3 \%$,
and their spin is either $0$ (preferably) or $1/2$.

Finally, other leptons and hadrons probably also have the texture 
structure in the vicinity of surface region whereas their differences are 
caused by different interiors and/or dynamical properties of
surface region as whole.

Shortly, in the spirit of the electromagnetic mass theory 
one can suggest that mass
is determined by charge because mass appears as the result of interaction 
of (charged and coupled) topological defects with space-time metric 
provided quantum uncertainty, therefore, charge is more 
fundamental property than mass.
In this connection the above-mentioned unification of particle families
should not to be mixed up
with the Standard Model one since it is based on the 
spatio-temporal and topological (i.e., pre-SM) assumptions.

\section{Conclusion}

Let us summarize briefly the main results obtained.
In Sec. \ref{s-in} we analyzed the existing extended particle models based 
on the discontinuities of first kind (boundary surfaces).
We pointed out that conditions of stability of such configurations require
that both the charge and internal vacuum energy are transferred to the 
configuration boundary.
This circumstance has to be in contradiction with the definition of the 
boundary surfaces as such and compels to consider the 
discontinuities of second kind (singular hypersurfaces) which can have the
surface charge, tension, and energy.
In Sec. \ref{s-si} such a model was constructed, thereby, following
aforesaid, we studied only hollow configurations. 
We obtained that the  conditions of equilibrium of the model require
the matter on a shell's surface to be of the global texture type.
Thereby we obtained that the charged texture in equilibrium
(which being neutral has a zero gravitational mass)
receives the mass due to the electromagnetic interaction.
In Sec. \ref{s-qu} we studied quantum aspects of the theory.
We constructed wave equations and obtained total mass (energy) spectra for 
bound states.
The most realistic spectrum we obtained contains two mass-energy
continua, 
thereby the Dirac sea conception arises in a natural way. 
Besides, in a natural way we got also a phenomenological unification of 
the two families of elementary particles, $\pi^\pm$-mesons and electrons 
and positrons, 
thereby the numerical value of the ratio of their masses was explained
within the frameworks of a quantum gravitational model.
Considering these facts we can suppose 
that the texture matter is probably another kind of 
fundamental matter both considering microscopical (quantum) properties 
and non-negligibly interacting with own gravitational field.
Following some modern cosmological ideas \cite{tur,vil} if texture matter 
exists then the Universe could be closed and still have a non-relativistic
matter density.
In this connection our model proposes the answer on the important question,
what are the real sources of texture matter.
Besides, the texture hypothesis means that in the early Universe before
lepton-hadron phase there existed the topological texture foam phase 
sufficiently affecting both on the modern observable features of particles
in our Universe and on evolution of the Universe as whole.
Besides, the texture foam can be released again in present epoch, e.g.,
inside energetically supersaturated astronomical objects 
where high temperature favours
to decay of the known particles into more elementary components.

Finally, the separately long discussed problem whether the additional quantum 
$\zeta$-splitting has any physical sense \cite{hkk}
was resolved in terms of two classes of elementary particles.

Thus, in the present paper we have consecutively constructed the 
classical and quantum  models of an extended 
charged particle based on the (2+1)-dimensional spherical hollow charged 
texture in general relativity.

\def\CMPh{Commun. Math. Phys.}
\def\JPh{J. Phys.}
\def\CJP{Czech. J. Phys.}
\def\FP{Fortschr. Phys.}
\def\LMPh {Lett. Math. Phys.}
\def\MPL {Mod. Phys. Lett.}
\def\NPh  {Nucl. Phys.}
\def\PhE  {Phys.Essays}
\def\PhL  {Phys. Lett.}
\def\PhR  {Phys. Rev.}
\def\PhRL {Phys. Rev. Lett.}
\def\PhRp {Phys. Rep.}
\def\NCim {Nuovo Cimento}
\def\NuPB {Nucl. Phys.}
\def\GRG {Gen. Relativ. Gravit.}
\def\CQG {Class. Quantum Grav.}
\def\prp {report}
\def\Prp {Report}

\def\jn#1#2#3#4#5{{\it #1}{#2} {\bf #3}, {#4} {(#5)}}
% #1 tittle  #2 ser  #3 vol  #4 page  #5 year
\def\boo#1#2#3#4#5{{\it #1} ({#2}, {#3}, {#4}){#5}}
% #1 tittle  #2 publisher  #3 place  #4 year  #5 page/, p.789/
\def\prpr#1#2#3#4#5{{``#1,''} {#2}{ #3}{ No. #4}, {#5} (unpublished)}
% #1 tittle  #2 place #3-P/preprint- #4No #5 year 


\begin{thebibliography}{99}

\bibitem{roh}
F. Rohrlich,
\boo{Classical charged particles}{Addison-Wesley}{Massachusetts}{1965}{}.

\bibitem{poi}
H. \Po,
\jn{C. R. Acad. Sci.}{}{140}{1504}{1905};
\jn{Rend. Circ. Mat. Palermo}{}{21}{129}{1906}.

\bibitem{gro}
\O. Gr\o n,
\jn{\PhR}{ D}{31}{2129}{1985}.

\bibitem{il}
W. Israel,
\jn{\PhR}{ D}{2}{641}{1970};
C. A. L\'opez,
\jn{\PhR}{ D}{30}{313}{1984}.

\bibitem{zlo008}
K. G. Zloshchastiev,
\jn{\CQG}{}{16}{1737}{1999}.


\bibitem{gau}
R. Gautreau,
\jn{\PhR}{ D}{31}{1860}{1985}

\bibitem{cc}
J. M. Cohen and M. D. Cohen,
\jn{\NCim}{ B}{60}{241}{1969}.

\bibitem{zel}
Ya. B. Zel'dovich,
\jn{Sov. Phys. Usp.}{}{11}{381}{1968}.

\bibitem{trk}
R. N. Tiwari, J. R. Rao, and R. R. Kanakamedala,
\jn{\PhR}{ D}{30}{489}{1984}.

\bibitem{col}
S. Coleman,
in
\boo{The Whys of Sub-Nuclear Physics}{Plenum-Press}{New York}{1979}{}.

\bibitem{dau}
G. Dautcourt, \jn{Math. Nachr.}{}{27}{277}{1964}.

\bibitem{isr}
W. Israel,
\jn{\NCim}{ B}{44}{1}{1966}.

\bibitem{mtw}
C. W. Misner, K. S. Thorne, and J. A. Wheeler, 
\boo{Gravitation}{Freeman}{San Francisco}{1973}{};
C. Barrab\'es and G. F. Bressange,
\jn{\CQG}{}{14}{805}{1997}.

\bibitem{fhk}
J. Frauendiener, C. Hoenselaers, and W. Konrad,
\jn{\CQG}{}{7}{585}{1990};
P. R. Brady, J. Louko, and E. Poisson,
\jn{\PhR}{ D}{44}{1891}{1991}.

\bibitem{tur}
R. L. Davis,
\jn{\PhR}{ D}{35}{3705}{1987};
N. G. Turok,
\jn{\PhRL}{}{63}{2625}{1989};
N. G. Turok and D. Spergel,
\jn{\PhRL}{}{64}{2736}{1990};
D. N\"otzold,
\jn{\PhR}{ D}{43}{R961}{1991};
K. G. Zloshchastiev,
\jn{\MPL}{ A}{13}{1419}{1998}.

\bibitem{dad}
N. Dadhich, 
\prpr{A Duality Relation: 
Global Monopole and Texture}{LANL}{\Prp}{gr-qc/9712021}{1997}.

\bibitem{vil} 
A. Vilenkin, 
\jn{\PhRL}{}{53}{1016}{1984};
R. J. Gott III and M. J. Rees, 
\jn{Mon. Not. R. Astr. Soc.}{}{227}{453}{1987};
E. W. Kobl, 
\jn{Ap. J.}{}{334}{543}{1989};
R. G. Crittenden and N. G. Turok,
\jn{\PhRL}{}{75}{2642}{1995};
M. Kamionkowski and N. Toumbas,
\jn{\PhRL}{}{77}{587}{1996}.

\bibitem{cg}
Y. Kim, K. Maeda, and N. Sakai,
\jn{\NPh}{ B}{481}{453}{1996};
I. Cho and J. Guven, 
\jn{\PhR}{ D}{58}{063502}{1998}.


\bibitem{zlo}
K. G. Zloshchastiev,
\jn{\PhR}{ D}{57}{4812}{1998}.

\bibitem{hb}
P. H\'aj\'\i\v{c}ek and J. Bi\v{c}\'{a}k, 
\jn{\PhR}{ D}{56}{4706}{1997};
J. L. Friedman, J. Louko, and S. N. Winters-Hilt, 
\jn{\PhR}{ D}{56}{7674}{1997}.

\bibitem{hkk}
P. H\'aj\'\i\v{c}ek, B. S. Kay, and K. V. Kucha\v{r}, 
\jn{\PhR}{ D}{46}{5439}{1992}; 
E. Hawkins, \jn{\PhR}{ D}{49}{6556}{1994}.

\bibitem{bkkt}
V. A. Berezin, N. G. Kozimirov, V. A. Kuzmin, and I. I. Tkachev, 
\jn{\PhL}{ B}{212}{415}{1988}; 
V. A. Berezin, \jn{\PhL}{ B}{241}{194}{1990}; 
\jn{Nucl. Phys. Proc. Suppl.}{}{57}{181}{1997}. 

\bibitem{not}
K. Nakamura, Y. Oshiro, and A. Tomimatsu, 
\jn{\PhR}{ D}{54}{4356}{1996}. 

\bibitem{vil2} 
A. Vilenkin,
\jn{\PhR}{ D}{50}{2581}{1994}.

\bibitem{gz}
V. D. Gladush and K. G. Zloshchastiev, unpublished.

\bibitem{bd}
J. D. Bjorken and S. D. Drell, \boo{Relativistic Quantum 
Mechanics}{McGraw-Hill}{NY}{1964}{};
L. D. Landau and E. M. Lifshitz, \boo{Course of Theoretical 
Physics}{Pergamon}{Oxford}{1971}{, Vol. IV}.

\bibitem{as}
{\it Handbook of Mathematical Functions}, edited by M. A. Abramowitz 
and I. A. Stegun (Dover, NY, 1972).

\end{thebibliography}
\end{document}